\documentclass[12pt]{article}

\usepackage{graphicx}

\begin{document}

\title{Snyder and their representation with creation and annihilation operators}

\author{Juan M. Romero \thanks{jromero@cua.uam.mx}\\[0.5cm]
\it Departamento de Matem\'aticas Aplicadas y Sistemas,\\
\it Universidad Aut\'onoma Metropolitana-Cuajimalpa,\\
\it  M\'exico, D.F  05300, M\'exico.\\
[0.5cm]\\
Leonardo Ort\'iz \thanks{leonardoortizh@ciencias.unam.mx}\\[0.5cm]
\it Departamento de F\'isica, Facultad de Ciencias,\\
\it Universidad Nacional Aut\'onoma de M\'exico,\\
\it A. Postal 04510, CDMX, M\'exico.
}

\date{}

\pagestyle{plain}

\maketitle

\begin{abstract}
Inspired by the Schwinger's representation of angular momentum, we propose a representation of certain operators where we use the algebra of the annihilation and creation operators. In particular, we propose a representation of the Snyder space-time with the help of the annihilation and creation operators, which create and annihilate quantum of space. In addition, we show that by using a matrix representation of the $SO(3)$ or $SU(2)$ Lie algebra it is possible to obtain a representation of the spacial sector of Snyder space-time. Finally, we obtain a quantized expectation value of the area of a sphere.\\

\end{abstract}

\section{Introduction}
The finding a quantum theory of gravity remains as a very important open problem in theoretical physics. With regarding to thisquestion there has been proposed several approaches such as string theory \cite{polchinski}, loop quantum gravity \cite{rovelli}, non-commutative geometry \cite{connes} and  Horava-Lifshitz gravity \cite{horava}. Although we do not have a theory of quantum gravity, from several lines of thought there are indications that there should be a quantization of geometric quantities. In this respect,  based on the Ehrenfest principle, Bekenstein's conjecture establishes the quantization of the area of the horizon of a black hole as \cite{beke1,beke2}
\begin{equation}
A_{n}=4\pi r^2=\gamma l_{p}^2n,\hspace{1cm}n=1, 2, \cdots, \nonumber
\end{equation}
where $\gamma$ is a proportionality constant and $l_{p}$ is
the Planck's length. Notice that $A_{n}$ represents quantums of the area of sphere of radius $r.$ Additionally, in $(2+1)$ dimensions there is available already a quantum theory of gravity \cite{w1,carlip}. Notably, in these dimensions, within an effective approach there has been obtained by  G. `t Hooft a non-commutative space well known and so-called Snyder space-time \cite{hooft}.  Which is discrete, non commutative and compatible with the Lorentz symmetry \cite{s}. Moreover, in this space the quantization of the area of a sphere has been carried out in a natural way \cite{za}. It is worth mentioning that it is possible to construct field theories in some non-commutative space-times \cite{w2,douglas,szabo,kontsevich}, but to build a gravity  or field theory in a non-commutative space-time, as Snyder space-time, is a very difficult task and remains as an unsolved problem. Some work about Snyder space-time can be seen in \cite{snyder1,snyder2,snyder3,snyder4,snyder5,snyder6,snyder7,snyder8,snyder9,snyder10,snyder11,snyder12,snyder13,snyder14} and  works about non-commutative space-times which imply discrete geometric quantities can be seen in \cite{dis1,dis2,dis3,dis4,dis5}. \\  

In this paper, inspired by a seminal work of Schwinger on angular momentum \cite{sch}, by using the annihilation and creation operators we propose a representation of a certain operators which satisfy different commutation relations. In particular, we propose a representation of the spacial sector of Snyder space-time using  annihilation and creation operators which create and annihilate quantum of space. This results allows us to obtain a quantized expectation value of the area of a sphere which is consistent with Bekenstein's conjecture. It is worth mentioning that this procedure can be done using a matrix representation of the $SO(3)$ or $SU(2)$ Lie algebra.\\

This paper is organized as follow: in section $2,$ by using the annihilation and creation operators, we propose a representation of a certain operators; in  section $3,$ we study the spacial sector of the Snyder space-time; in  section $4,$ by using a matrix representation of $SO(3)$ and annihilation and creation operators we obtain a representation of the spacial coordinates of Snyder space-time; in section $5,$ by using a matrix representation of $SU(2)$ and annihilation and creation operators we obtain a representation of the spacial coordinates of Snyder space-time; in section $6,$ a summary is given.

 \section{Lie algebras and their representation with creation and annihilation operators}
 
 Suppose that 
 \begin{eqnarray}
 C_{\mu},  \qquad  \mu=0,1,2, \cdots d, 
 \end{eqnarray}
 are hermitian matrices of $n\times n,$ which satisfy the commutation relations    
 \begin{eqnarray}
 [C_{\mu},C_{\nu}]= B_{\mu\nu}\nonumber  
 \end{eqnarray}
 where 
 \begin{eqnarray}
 B_{\mu\nu}\nonumber  
 \end{eqnarray}
 is an hermitian antisymmetric tensor. \\
 
In addition suppose that 
 \begin{eqnarray}
\hat a_{i},\qquad   i=,1,2, \cdots n, \label{oca}
 \end{eqnarray}
 are $n$ operators which satisfy the commutation relations    
\begin{eqnarray}
\left[\hat a_{k},\hat a_{l}\right]&=&0\\
\left[\hat a_{k},\hat a^{\dagger}_{l}\right]&=&\delta_{kl}\\
\left[\hat a^{\dagger}_{k},\hat a^{\dagger}_{l}\right]&=&0. \nonumber
\end{eqnarray}
Notice that using the operators (\ref{oca}) we can construct the operators  
\begin{eqnarray}
\hat {\cal  C}_{\mu}=C_{\mu kl}\hat a^{\dagger}_{k}\hat a_{l}, \nonumber
\end{eqnarray}
\begin{eqnarray}
 \hat {\cal B}_{\mu\nu}=B_{\mu\nu kl}\hat a^{\dagger}_{k}\hat a_{l}.\nonumber
\end{eqnarray}
In this case we obtain
\begin{eqnarray}
\left[ \hat {\cal C}_{\mu}, \hat {\cal C}_{\nu}\right]&=&\left[C_{\mu kl}\hat a^{\dagger}_{k}\hat a_{l}, C_{\nu rs}\hat a^{\dagger}_{r}\hat a_{s}\right]= C_{\mu kl}C_{\nu rs}\left[\hat a^{\dagger}_{k}\hat a_{l}, \hat a^{\dagger}_{r}\hat a_{s}\right]\nonumber\\
&=& C_{\mu kl}C_{\nu rs}\left( \hat a^{\dagger}_{k}\left[ \hat a_{l}, \hat a^{\dagger}_{r}\right] \hat a_{s}+\hat a^{\dagger}_{r} \left[\hat a^{\dagger}_{k}, \hat a_{s}\right] \hat a_{l}\right)\nonumber\\
&=& C_{\mu kl}C_{\nu rs}\left( \hat a^{\dagger}_{k} \hat a_{s}\delta_{lr} -\hat a^{\dagger}_{r} \hat a_{l}\delta_{sk} \right)\nonumber\\
&=& C_{\mu kr}C_{\nu rs} \hat a^{\dagger}_{k} \hat a_{s}-C_{\mu kl}C_{\nu rk}\hat a^{\dagger}_{r} \hat a_{l}\nonumber\\
&=& C_{\mu kr}C_{\nu rs} \hat a^{\dagger}_{k} \hat a_{s}-C_{\mu rs}C_{\nu kr}\hat a^{\dagger}_{k} \hat a_{s}\nonumber\\
&=&\left (C_{\mu kr}C_{\nu rs} -C_{\mu rs}C_{\nu kr}\right) \hat a^{\dagger}_{k} \hat a_{s}\nonumber\\
&=&\left (C_{\mu kr}C_{\nu rs} -C_{\nu kr}C_{\mu rs}\right) \hat a^{\dagger}_{k} \hat a_{s}\nonumber\\
&=&\left (C_{\mu}C_{\nu} -C_{\nu}C_{\mu }\right)_{ks} \hat a^{\dagger}_{k} \hat a_{s}\nonumber\\
&=&\left [C_{\mu}, C_{\nu} \right]_{ks} \hat a^{\dagger}_{k} \hat a_{s}\nonumber\\
&=&\left(B_{\mu\nu}\right)_{ks}\hat a^{\dagger}_{k} \hat a_{s},\nonumber
\end{eqnarray}
namely
\begin{eqnarray}
\left[ \hat {\cal C}_{\mu}, \hat {\cal C}_{\nu}\right]&=&\hat {\cal B}_{\mu\nu}.\nonumber
\end{eqnarray}
In the next sections this result will be employed.

\section{Snyder space-time} 
 
 The commutation relations of 
 the non-commutative Snyder space-time are given by  \cite{s}:
\begin{eqnarray}
\left [ \hat x^{\mu}, \hat x^{\nu} \right ] &=&-i l^{2}  \left( \hat x^{\mu} \hat p^{\nu} - \hat x^{\nu} \hat p^{\mu} \right), \label{s-cr1} \\
\left[ \hat x^{\mu}, \hat p^{\mu} \right]&=&i \left( \eta^{\mu \nu} - l^{2} \hat p^{\mu}\hat p^{\nu}\right),\label{s-cr2}\\
\left[ \hat p^{\mu}, \hat p^{\mu} \right]&=&0. \label{s-cr3}
\end{eqnarray}
where $l$ is a constant. Notice that the operators 
\begin{eqnarray}
\hat x^{\mu}&=&-il\left(\zeta\frac{\partial }{\partial \zeta_{\mu}}-\zeta^{\mu}\frac{\partial }{\partial \zeta }\right),\nonumber\\
\hat p_{\mu}&=& -\frac{1}{l}\frac{\zeta_{\mu }}{\zeta }, \nonumber
\end{eqnarray}
 satisfy the commutation relations (\ref{s-cr1})-(\ref{s-cr3}).\\
 
 In addition, it can be shown that the spacial sector of the commutation relations (\ref{s-cr1}) can be written as 
\begin{eqnarray}
\left [ \hat x_{i}, \hat x_{j} \right ] &=&i l  \epsilon_{ijk} \hat {\cal L}_{k},\nonumber
\end{eqnarray}
 where
\begin{eqnarray}
\hat  {\cal L}_{i}=-il\epsilon_{ijk}  \zeta^{j}\frac{\partial }{\partial \zeta^{k}}. \nonumber
\end{eqnarray}
In addition it can be seen that the  operators $ \hat {\cal L}_{i}$ satisfy 
\begin{eqnarray}
\left [ \hat {\cal L}_{i}, \hat {\cal L}_{j} \right ] &=&i l  \epsilon_{ijk}{\cal L}_{k},\nonumber\\
\left [ \hat x_{i}, \hat {\cal L}_{j} \right ] &=&i l  \epsilon_{ijk}x_{k}.\nonumber
\end{eqnarray}

Now, lets us to introduce the following operators 
\begin{eqnarray}
\hat M_{i}&=&\frac{1}{2}\left( \hat {\cal L}_{i}+\hat x_{i}  \right), \nonumber\\
\hat N_{i}&=&\frac{1}{2}\left(  \hat {\cal L}_{i}- \hat x_{i}  \right), \nonumber
\end{eqnarray}
which satisfy 
\begin{eqnarray}
\left [ \hat M_{i}, \hat M_{j} \right ] &=&i l,  \epsilon_{ijk}M_{k},\label{so1}\\
\left [ \hat N_{i}, \hat N_{j} \right ] &=&i l,  \epsilon_{ijk}N_{k},\label{so2}\\
\left [ \hat N_{i}, \hat M_{j} \right ] &=&0.  \label{so3}
\end{eqnarray}
Notice that the commutation relations (\ref{so1})-(\ref{so3}) can be satisfied by the generators of $SO(3)$ or $SU(2)$ group. Below, in the following sections, we will study both cases.

\section{$SO(3)$ case}

We can see that the matrices
\begin{eqnarray}
m_{1}=il
\left(
\begin{array}{rrr}
 0& 0&0 \\
 0& 0 &-1\\
 0&1&0 
\end{array}\right),
m_{2}=il
\left(
\begin{array}{rrr}
 0& 0&0 \\
 0& 0 &-1\\
 0&1&0 
\end{array}\right),
m_{3}=il
\left(
\begin{array}{rrr}
 0& -1&0 \\
 1& 0 &0\\
 0&0&0 
\end{array}\right),
\nonumber
\end{eqnarray}
satisfy the commutation relations
\begin{eqnarray}
\left [ m_{i},  m_{j} \right ] &=&i l  \epsilon_{ijk}m_{k}.\nonumber
\end{eqnarray}
Then, it can be proposed the following operators 
\begin{eqnarray}
\hat M_{i}&=&\left(
\begin{array}{rrr}
 \hat a_{1}^{\dagger}&
 \hat a_{2}^{\dagger}&
 \hat a_{3}^{\dagger}
\end{array}\right) m_{i}\left(
\begin{array}{r}
 \hat a_{1}\\
 \hat a_{2}\\
 \hat a_{3}
\end{array}\right)\nonumber\\
\hat N_{i}&=&\left(
\begin{array}{rrr}
 \hat b_{1}^{\dagger}&
 \hat b_{2}^{\dagger}&
 \hat b_{3}^{\dagger}
\end{array}\right) m_{i}\left(
\begin{array}{r}
 \hat b_{1}\\
 \hat b_{2}\\
 \hat b_{3}
\end{array}\right),\nonumber
\end{eqnarray}
which satisfy the commutation relations (\ref{so1})-(\ref{so3}) .\\ 

Notice that these operators can be written as
\begin{eqnarray}
\hat M_{1}&=&il\left(\hat a^{\dagger}_{2}\hat a_{3}-\hat a^{\dagger}_{3}\hat a_{2} \right),\hat M_{2}=il\left(\hat a^{\dagger}_{3}\hat a_{1}-\hat a^{\dagger}_{1}\hat a_{3} \right), \hat M_{3}=il\left(\hat a^{\dagger}_{1}\hat a_{2}-\hat a^{\dagger}_{2}\hat a_{3} \right),\nonumber\\
\hat N_{1}&=&il\left(\hat b^{\dagger}_{2}\hat b_{3}-\hat b^{\dagger}_{3}\hat b_{2} \right),\hat N_{2}=il\left(\hat b^{\dagger}_{3}\hat b_{1}-\hat b^{\dagger}_{1}\hat b_{3} \right), \hat N_{3}=il\left(\hat b^{\dagger}_{1}\hat b_{2}-\hat b^{\dagger}_{2}\hat b_{3} \right).\nonumber
\end{eqnarray}
Now because
\begin{eqnarray}
\hat x_{i}=\hat M_{i}-\hat N_{i},\nonumber
\end{eqnarray}
it can be obtained
\begin{eqnarray}
\hat x_{1}&=&il\left(\hat a^{\dagger}_{2}\hat a_{3}-\hat a^{\dagger}_{3}\hat a_{2}-\hat b^{\dagger}_{2}\hat b_{3}+\hat b^{\dagger}_{3}\hat b_{2} \right)\label{p-sn1}\\
\hat x_{2}&=&il\left(\hat a^{\dagger}_{3}\hat a_{1}-\hat a^{\dagger}_{1}\hat a_{3}-\hat b^{\dagger}_{3}\hat b_{1}+\hat b^{\dagger}_{1}\hat b_{3} \right),\\ 
\hat x_{3}&=&il\left(\hat a^{\dagger}_{1}\hat a_{2}-\hat a^{\dagger}_{2}\hat a_{1} -\hat b^{\dagger}_{1}\hat b_{2}+\hat b^{\dagger}_{2}\hat b_{1}\right).\label{p-sn3}
\end{eqnarray}
Notice that these expression indicate that the creation and annihilation operators $\hat a_{i},\hat a^{\dagger}_{i},\hat b_{i},\hat b^{\dagger}_{i}, (i=1,2,3)$ create and annihilate quantum of space.\\

Now, using the equation (\ref{p-sn1})-(\ref{p-sn3}) it can be obtained the operators 
\begin{eqnarray}
\hat r^{2}&=&\hat x^{2}_{1}+\hat x^{2}_{2}+\hat x^{2}_{3}\nonumber\\
&=&l^{2}\Bigg(\hat a^{\dagger}\cdot \hat a+ \left(\hat a^{\dagger}\cdot \hat a\right)^{2}- \left(\hat a^{\dagger} \right)^{2} \left(\hat a \right)^{2}+ \hat b^{\dagger}\cdot \hat b+ \left(\hat b^{\dagger}\cdot \hat b\right)^{2}- \left(\hat b^{\dagger} \right)^{2} \left(\hat b \right)^{2}\nonumber\\
& & +2\left(
\left(\hat a^{\dagger}\cdot \hat b\right)
\left(\hat b^{\dagger}\cdot \hat a\right)-
\left(\hat a^{\dagger}\cdot \hat b^{\dagger}\right)
\left(\hat a\cdot \hat b\right) \right)
\Bigg)
\end{eqnarray}
Using this expression, it can be constructed the operator of sphere area as
\begin{eqnarray}
\hat A=4 \pi \hat r^{2}, \label{area-s}
\end{eqnarray}
whose expectation value  is 
\begin{eqnarray}
& &\left< n_{1},n_{2},n_{3},m_{1},m_{2},m_{3}|\hat A|m_{1},m_{2},m_{3},n_{1},n_{2},n_{3}\right>\nonumber\\
& &=8l^{2} \pi\Bigg( n_{1}+n_{2}+n_{3}+m_{1}+m_{2}+m_{3} 
+n_{1}n_{2}+n_{3}n_{1}+n_{3}n_{2}\nonumber\\
& &+m_{1}m_{2}+m_{1}m_{3}+m_{2}m_{3}\Bigg). \label{area-q}
\end{eqnarray}
Thus, we obtain a quantized expectation value of the area of a sphere.\\ 

Now, it can be shown that in the Schwarzschild space-time the area of a sphere is given by $A=4\pi r^{2},$ then the quantum version of this geometric quantity is (\ref{area-s}). Thus, it can be seen that the expectation value  (\ref{area-q}) is consistent with the Bekenstein's conjecture.

\section{$SU(2)$ case}

Now, notice that the matrices
\begin{eqnarray}
m_{1}=\frac{l}{2}
\left(
\begin{array}{rrr}
 0& 1 \\
 1& 0 
\end{array}\right),
m_{2}=i\frac{l}{2}
\left(
\begin{array}{rrr}
 0& -1\\
 1&0 
\end{array}\right),
m_{3}=\frac{l}{2}
\left(
\begin{array}{rrr}
 1&0 \\
 0 &-1
\end{array}\right),
\nonumber
\end{eqnarray}
satisfy the commutation relations
\begin{eqnarray}
\left [ m_{i},  m_{j} \right ] &=&i l  \epsilon_{ijk}m_{k}.\nonumber
\end{eqnarray}
Then, it can be proposed the following operators 
\begin{eqnarray}
\hat M_{i}&=&\left(
\begin{array}{rrr}
 \hat a_{1}^{\dagger}&
 \hat a_{2}^{\dagger}
\end{array}\right) m_{i}\left(
\begin{array}{r}
 \hat a_{1}\\
 \hat a_{2}
\end{array}\right)\label{n1}\\
\hat N_{i}&=&\left(
\begin{array}{rrr}
 \hat a_{3}^{\dagger}&
 \hat a_{4}^{\dagger}
\end{array}\right) m_{i}\left(
\begin{array}{r}
 \hat a_{3}\\
 \hat a_{4}
\end{array}\right),\label{n2}
\end{eqnarray}
which satisfy the commutation relations (\ref{so1})-(\ref{so3}).\\ 

Now, by using the operators (\ref{n1})-(\ref{n2}), it can be obtained 
\begin{eqnarray}
\hat x_{1}&=&\frac{l}{2}\left( a^{\dagger}_{1}a_{2}-a^{\dagger}_{3}a_{4}+a^{\dagger}_{2}a_{1}-a^{\dagger}_{4}a_{3} \right)\label{p2-sn1}\\
\hat x_{2}&=&\frac{l}{2i}\left( a^{\dagger}_{1}a_{2}-a^{\dagger}_{3}a_{4}-a^{\dagger}_{2}a_{1}+a^{\dagger}_{4}a_{3} \right)\\
\hat x_{3}&=&\frac{l}{2}\left(a^{\dagger}_{1}a_{1}-a^{\dagger}_{2}a_{2}-a^{\dagger}_{3}a_{3}+a^{\dagger}_{4}a_{4} \right).\label{p2-sn3}
\end{eqnarray}
Then, the creation and annihilation operators $\hat a_{i},\hat a^{\dagger}_{i}, (i=1,2,3,4)$ create and annihilate quantum of space.\\

Now, using the operators (\ref{p2-sn1})-(\ref{p2-sn3}) it can be obtained
\begin{eqnarray}
< n_{1},n_{2}, n_{2}, n_{4}|\hat x_{1} | n_{1},n_{2}, n_{2}, n_{4}>&=&0,\nonumber\\
< n_{1},n_{2}, n_{2}, n_{4}|\hat x_{2} | n_{1},n_{2}, n_{2}, n_{4}>&=&0,\nonumber\\
< n_{1},n_{2}, n_{2}, n_{4}|\hat x_{3} | n_{1},n_{2}, n_{2}, n_{4}>&=&\frac{l}{2}\left(n_{1}+n_{4}-\left(n_{2}+n_{3}\right) \right).\nonumber
\end{eqnarray}
Moreover, it can be constructed the following operator
\begin{eqnarray}
\hat r^{2}= \hat x_{1}^{2}+\hat x_{2}^{2}+\hat x_{3}^{2}&=&\frac{l^{2}}{4}\Bigg(  \hat n^{2}_{3}+\hat n^{2}_{4} +\hat n^{2}_{1}+\hat n^{2}_{2}+ 2 \left(\hat n_{1}+\hat n_{2} +\hat n_{3}+\hat n_{4} \right) \nonumber\\
& &+ 2\hat n_{3}\left(\hat n_{4}+\hat n_{2}-\hat n_{1} \right)+2\hat n_{4}\left(\hat n_{1}-\hat n_{2}\right) +2\hat n_{1}\hat n_{2}\nonumber\\
& &-2\left(\hat a^{\dagger}_{2}\hat a^{\dagger }_{3}\hat a_{4}\hat a_{1}+\hat a^{\dagger}_{4}\hat a^{\dagger}_{1}\hat a_{2}\hat a_{3}+ \hat a^{\dagger}_{3}\hat a^{\dagger}_{1}\hat a_{4}\hat a_{2}+\hat a^{\dagger}_{4}\hat a^{\dagger}_{2}\hat a_{3}\hat a_{1} \right) \Bigg). \nonumber
\end{eqnarray}
Then for the area operator of the sphere  it can be obtained
\begin{eqnarray}
\left< n_{1},n_{2}, n_{2}, n_{4}|\hat A| n_{1},n_{2}, n_{2}, n_{4}\right>&=&\left< n_{1},n_{2}, n_{2}, n_{4}|4\pi \hat r^{2}| n_{1},n_{2}, n_{2}, n_{4}\right>\nonumber\\
&= &l^{2}\pi \Bigg(   n^{2}_{1}+ n^{2}_{2} + n^{2}_{3}+ n^{2}_{4}+ 2 \left( n_{1}+n_{2} + n_{3}+ n_{4} \right) \nonumber\\
& &+ 2n_{1}\left( n_{2}+ n_{4}- n_{3} \right)+2 n_{2}\left( n_{3}-n_{4}\right) +2n_{3} n_{4} \Bigg)\nonumber
\end{eqnarray}
namely 
\begin{eqnarray}
\left< \hat A\right>&=&l^{2}\pi \Bigg(   \left( n_{3}-n_{1}\right)^{2}+\left( n_{4}-n_{2}\right)^{2}+ 2 \left( n_{1}+n_{2} + n_{3}+ n_{4} \right)\nonumber\\
& &+ 2n_{3}\left( n_{2}+ n_{4} \right)+2 n_{4}n_{1} +2n_{1} n_{2} \Bigg),\nonumber
\end{eqnarray}
notice that when 
\begin{eqnarray}
n_{1}=n_{3}, \quad n_{2}=n_{4} \nonumber
\end{eqnarray}
it can be obtained
\begin{eqnarray}
\left< \hat A\right>&=&4\pi l^{2} \left(   n_{1}+n_{2} +4n_{1}n_{2}\right). \nonumber
\end{eqnarray}
This result is also consistent with the Bekenstein's conjecture.

\section{Summary}
\label{summary-sec}
In this article, inspired by a seminal work of Schwinger on angular momentum \cite{sch}, we proposed a representation of the spatial sector of Snyder space-time using  annihilation and creation operators, which create and annihilate quantum of space. And it allowed us to obtain a quantized expectation value of the area of a sphere. In order to obtain these results we employed the $SO(3)$ or $SU(2)$ Lie algebra.\\

It is worth mentioning that with the annihilation and creation operators coherent states can be obtained. Then  
in a future work, we plan study coherent states for the Snyder space-time. In particular we shall study the Snyder version of the Fourier transform as in the reference \cite{espa} where it has been done in the standard non-commutative space. Armed with this tool, the Fourier transform for the Snyder space-time, we will address the corrections to some black hole metrics as has been done in \cite{nico} for the standard constant non-commutative space.     

 \section*{Acknowledgments} 
The authors thank SNI for support.

\end{document}